\documentclass[aps,showpacsx,twocolumn]{revtex4-1}
\usepackage{graphicx}
\usepackage{bm}
\usepackage{diagbox}
\usepackage{amsmath}
\usepackage{array}
\usepackage{color}
\usepackage{eufrak}
\usepackage{mathrsfs}
\usepackage{ulem}

\begin{document}
\title{Fermion spectral function in hot strongly interacting matter from the functional renormalization group}

\author{Ziyue Wang$^{1}$}
\email{zy-wa14@mails.tsinghua.edu.cn}

\author{Lianyi He$^{1,2}$}

\affiliation{$^{1}$ Physics Department, Tsinghua University, Beijing 100084, China}
\affiliation{$^{2}$ State Key Laboratory of Low-Dimensional Quantum Physics, Tsinghua University, Beijing 100084, China}

\date{\today}
\begin{abstract}
We present the first calculation of fermion spectral function at finite temperature in quark-meson model in the framework of the functional renormalization group (FRG). We compare the results in two truncations, after first evolving flow equation of effective potential, we investigate the spectral function either by taking the IR values as input to calculate one-loop self-energy or by taking the scale-dependent values as input to evolve the flow equation of the fermion two-point function. The latter one is a self-consistent procedure in the framework of FRG. In both truncations, we find a multi-peak structure in the spectral function, indicating quark collective excitations realized in terms of the Landau damping. However, in contrast to fermion zero-mode in the one-loop truncation, we find a fermion soft-mode in the self-consistent truncation, which approaches the zero-mode as temperature increases. 
\end{abstract}
\maketitle

\section{Introduction}
\label{s1}
The Quantum Chromodynamics (QCD) phase transitions at finite temperature and density provide a deep insight into the strong interacting matter created in high energy nuclear collisions and compact stars. The properties of the QGP phase near the critical temperature ($T_c$) acquire much interest, the heavy-ion collisions has suggested that the QGP matter is an ideal fluid \cite{Adcox:2004mh, Adams:2005dq}, indicating that the created matter is a strongly coupled system. The spectral properties of quark and hadron in this strongly interacting matter, are of fundamental importance for identifying the relevant degrees of freedom in the equation of state and respective transport properties. 

Whether quark can be described by well-defined quasi-particles has long been investigated, quasi-particles correspond to peaks with a small width in the spectral function with relevant quantum numbers. Quenched lattice QCD simulation indicates the existence of the quasi-particles of quarks with small decay width \cite{Karsch:2007wc,Karsch:2009tp,Kaczmarek:2012mb}. Finite temperature gauge theory with Dyson-Schwinger equation also predicts quasi-particle properties of quarks \cite{Harada:2008vk, Mueller:2010ah}. At high temperature, where the hard thermal loop (HTL) approximation applies, quarks still have some collective excitation known by the normal quasi-quark and the plasmino branches in the spectral function \cite{Pisarski:1988vd,Braaten:1989mz,Su:2014rma}. It has also been investigated that, quarks in the QGP phase can be described within a quasi-particle picture with a multi-peak spectral function~\cite{Kitazawa:2005mp,Kitazawa:2007ep,Kitazawa:2006zi}, whenever the interaction is mediated by scalar, pseudo-scalar, vector and axial-vector meson, which may exist as bosonic excitations in the QGP phase \cite{Kitazawa:2006zi}.

In the vicinity of $T_c$ of the chiral phase transition, non-perturbative effects are important, and one may expect, that the quark spectral functions will possess novel properties, when non-perturbative methods are adopted. In this paper, we employ the functional renormalization group (FRG)~\cite{Berges:2000ew,Polonyi:2001se,Pawlowski:2005xe,Gies:2006wv,Kopietz:2010zz, Braun:2011pp}  approach to the quark-meson model. As a non-perturbative method, FRG enables us to incorporate fluctuation effects beyond mean field theory, see Refs.~\cite{Berges:2000ew}. The self-consistent treatment of fluctuations is important towards the understanding of physics near a phase transition. Since the FRG allows a description of scale transformation, it provides a deep insight into the system where scale dependence plays a crucial role. To calculate the spectral function in the usually used imaginary time formalism with FRG, an analytical continuation is required to bring the imaginary time in the Euclidean two point function at finite temperature to the real time in the Minkowski space \cite{Pawlowski:2015mia,Strodthoff:2016pxx,Kamikado:2013sia,Tripolt:2013jra,Tripolt:2014wra,Jung:2016yxl,Tripolt:2016cey}. This method has been applied to the study of real time observables such as shear viscosity \cite{Tripolt:2016cey} and soft modes~\cite{Yokota:2016kyz} near the QCD critical point. The consistent investigation of spectral function in the framework of the FRG has been applied to various systems, including meson spectral function in a chiral phase transition~\cite{Tripolt:2013jra,Tripolt:2014wra}, quark spectral function in vacuum~\cite{Tripolt:2018qvi}, meson spectral function in a pion superfluid system~\cite{Wang:2017vis}.

When the FRG is put to use to investigate the quark spectral function, the most crucial difference is that one takes into account the scale dependence of the meson masses, and hence the thresholds of each decay, creation and scattering channel. The scale dependence is a high order effect and brings about difference in spectral function mainly in the following three aspects. First, it gives arise to novel structures in the imaginary part and real part of the self-energy. Second, the Landau damping which is the dominant effect at high temperature, is forbidden at low energy when considering the scale dependence of the meson masses, and leads to more peaks at low energy region at high temperature. Third, it is found that a fermion zero mode starts to appear when temperature is comparable to meson mass~\cite{Kitazawa:2006zi}, which also originates from the Landau damping effect. However, when the FRG is adopted, this zero mode becomes a soft mode, and approaches the origin when temperature increases.  

We organize the paper as follows. The FRG flow equations for the effective potential and the two truncations to calculate the quark self-energy are derived in Section \ref{s2}. The procedure to solve the flow equations and the numerical results are presented in Section \ref{s3}. We summarize in Section \ref{s4}.

\section{Flow equations and Truncation}
\label{s2}
As an low energy effective model, the quark-meson model comes from the partial bosonization of the four-fermion interaction model and exhibits many of the global symmetries of QCD. It is widely used as an effective chiral model to demonstrate the spontaneous chiral symmetry breaking in vacuum and its restoration at finite temperature and density~\cite{Jungnickel:1995fp,Schaefer:2004en, Pawlowski:2014zaa}. Here we take the two-flavor version of the model with pseudo-scalar mesons ${\bf \pi}$ and scalar meson $\sigma$ as the dominant meson degrees of freedom at energy scale up to $\Lambda\approx 1$ GeV. The Euclidean effective action of the model at finite temperature $T$ and density $\mu$ is given as
\begin{eqnarray}
\Gamma = \int_x&\Big[&\bar\psi\left(\partial\!\!\!/-\gamma_0\mu\right)\psi+g\bar\psi\left(\sigma+i\gamma_5{\vec \tau}\cdot{\vec \pi}\right)\psi\nonumber\\
&+&\frac{1}{2}(\partial_\mu\phi)^2+U(\phi^2)-c\sigma\Big],
\end{eqnarray}
where the abbreviation $\int_x$ stands for $\int_0^\beta dx_0\int d^3x$ with the inverse of temperature $\beta=1/T$, and ${\vec \tau}$ are the Pauli matrices in flavor space. The Yukawa coupling is chosen as $g=3.2$ to fit the quark mass in vacuum. The fermion field $\psi$ and meson field $\phi$ are defined as $\psi=(u, d)$ and $\phi=(\sigma,\pi_1,\pi_2,\pi_0)$. The explicit chiral symmetry breaking term $-c\sigma$ corresponds to a finite current quark mass $m_0$.

Quantum and thermal fluctuations are of particular importance in the vicinity of a phase transition and are conveniently included within the framework of FRG. The core quantity in this approach is the averaged effective action $\Gamma_k$ at the RG scale $k$ in Euclidean space, its scale dependence is described by the flow equation~\cite{Berges:2000ew,Polonyi:2001se,Pawlowski:2005xe,Gies:2006wv,Kopietz:2010zz, Braun:2011pp}
\begin{eqnarray}
\label{gamma}
\dot{\Gamma}_k =\text{Tr}\int_p\left[\frac{1}{2}G_{\phi,k}(p)\dot{R}_{\phi,k}(p)-G_{\psi,k}(p)\dot{R}_{\psi,k}(p)\right],
\end{eqnarray}
where $\dot{\Gamma}_k=\partial_k\Gamma_k$ and so as for $\dot{R}_k$. The symbol '$\text{Tr}$' represents the summation over all inner degrees of freedom of mesons and quarks.  
\begin{eqnarray}
\label{g}
G_{\phi,k}(q) &=& \left(\Gamma_k^{(2)}[\phi]+R_{\phi,k}(q)\right)^{-1},\nonumber\\
G_{\psi,k}(q) &=& \left(\Gamma_k^{(2)}[\psi]+R_{\psi,k}(q)\right)^{-1}
\end{eqnarray}
are the FRG modified meson and quark propagators with the two-point functions $\Gamma_k^{(2)}[\phi]=\delta^2\Gamma_k/\delta\phi^2$ and $\Gamma_k^{(2)}[\psi]=\delta^2\Gamma_k/\delta\psi\delta\bar\psi$ and the two regulators $R_{\phi,k}$ and $R_{\psi,k}$. The evolution of the flow from the ultraviolet limit $k=\Lambda$ to infrared limit $k=0$ encodes in principle all the quantum and thermal fluctuations in the action. To suppress the fluctuations with momentum smaller than the scale $k$ during the evolution, an infrared regulator $R$ is introduced in the flow equation. At finite temperature and density where the Lorentz symmetry is broken, we employ the optimized regulator function which is the three dimensional analogue of the 4-momentum regulator ~\cite{Litim:2000,Litim:2001up}. The bosonic and fermionic regulators are chosen to be
\begin{eqnarray}
R_{\phi,k}(p) &=& \vec p^2 r_B(y),\nonumber\\
R_{\psi,k}(p) &=& \vec \gamma\cdot \vec  p r_F(y)
\end{eqnarray}
in momentum space with $y=\vec  p^2/k^2$ and $r_B(y)=(1/y-1)\Theta(1-y)$ and $r_F(y)=(1/\sqrt y-1)\Theta(1-y)$. The regulators $R_{\phi,k}$ and $R_{\psi,k}$ in the propagators $G_\phi$ and $G_\psi$ amount to having regularized three-momenta $\vec  p_r^2=\vec  p^2(1+r_B(y))$ and $\vec p_r=\vec  p(1+r_F(y))$ for bosons and fermions respectively. The three dimensional regulators break down the Lorentz symmetry in vacuum. However, physical quantities are measured in the ground state at $k=0$, where the regulators vanish and the Lorentz symmetry is guaranteed.

In order to derive the meson and quark propagators $G_{\phi,k}$ and $G_{\psi,k}$, we expand the effective potential around the mean field $\langle\sigma\rangle_k$, which describes the chiral symmetry breaking, and introduce the chiral invariant $\rho_k=\langle\sigma\rangle_k^2$. The RG-modified propagators of mesons and quark are
\begin{eqnarray}
G_{\phi,k}^{-1}&=&p_0^2+\vec p_r^2+m_{\phi,k}^2,\nonumber\\
G_{\psi,k}^{-1}&=&-\gamma_0(ip_0+\mu)+\vec\gamma\cdot\vec p+m_{f,k},
\end{eqnarray}
with $m_{\sigma,k}^2=2U'+4\rho_k U''$ and $m_{\pi,k}^2=2U'$, $U'$ and $U''$ are first and second order derivatives of effective potential with respect to $\rho$, and quark mass $m_{f,k}=g\langle\sigma\rangle_k$.

Assuming uniform field configurations, the integral over space and imaginary time becomes trivial, and the effective action $\Gamma_k=\beta V U_k$ is fully controlled by the potential $U_k$ with $V$ and $\beta$ being the space and time regions of the system. The flow equation $U_k$ hence comes directly from that of $\Gamma_k$, namely $\partial_kU_k=(T/V)\partial_k\Gamma_k$. The flow equation of the effective potential is then calculated by
\begin{eqnarray}
\partial_kU_k~=~\frac{1}{2}J_\phi(E_{\sigma,k})+\frac{3}{2}J_\phi(E_{\pi,k})-N_cN_fJ_\psi(E_{\psi,k}),
\label{flowU}
\end{eqnarray}
with $J_\phi$ and $J_\psi$ are one-loop threshold functions, the explicit expressions are presented in Appendix by Eq.(\ref{A2}), and the energies are given by $E_{\phi,k}=\sqrt{k^2+m_{\phi,k}^2}$ and $E_{\psi,k}=\sqrt{k^2+m_{\psi,k}^2}$.

\begin{figure}[!hbt]\centering
\includegraphics[width=0.48\textwidth]{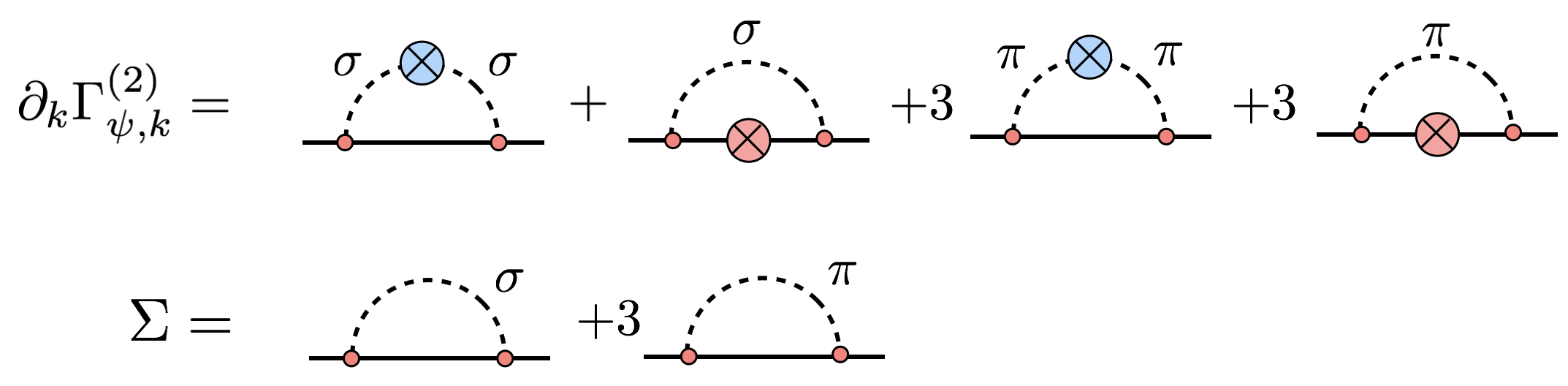}
\caption{The diagrammatic presentation of self-energy in two truncations, $\partial_k\Gamma_{\psi,k}^{(2)}$ is the flow equation for fermion two-point function in truncation A, $\Sigma$ is the one-loop self-energy in truncation B.}
\label{fig1}
\end{figure}

In the following, we are going to present two truncations to calculate the self-energy and spectral function. In both truncations, we first evolve the flow of the effective potential, and then take the masses as input to calculate the self-energy. In truncation A, we take scale-dependent masses $m_{\sigma,k}$ and $m_{\pi,k}$ at IR-minimum $\rho_{k=0}$ as input to integrate the flow of the two point function; while in truncation B, we take the IR masses $m_{\sigma,k=0}$ and $m_{\pi,k=0}$ to directly calculate the one-loop self-energy of quark. The diagrammatic description is presented in FIG.\ref{fig1}

\subsection{Truncation A}
In truncation A, we first evolve the flow of effective potential and prepare the scale-dependent meson masses as input. We then integrate the flow equation of the two-point function to obtain the self-energy in the infrared limit. The flow equation of fermion two-point function has Dirac structure, namely is a $4\times4$ matrix in Dirac space. For u quark of a certain color, the flow equation of the two-point function
\begin{eqnarray}
\partial_k{\Gamma}^{(2)}_{\bar{u}u}(p)~=~-g^2\widetilde{\partial}_k\int_q&\Big[&G_\sigma(q-p)G_u(q)\\
&+&3G_\pi(q-p)(i\gamma_5)G_u(q)(i\gamma_5)\Big],\nonumber
\end{eqnarray}
where $\widetilde{\partial}_k$ is the derivative of RG-scale k that only acts on the regulator $R_{\phi,k}$ and $R_{\psi,k}$ in the propagators. In this work, we consider only the spectral function at zero external momentum $\vec p=0$. In the ultraviolet, the Euclidean inverse quark propagator at $\vec{p}=0$ at the IR expansion point is 
\begin{eqnarray}
G^{-1}_{E,\Lambda}(ip_0)=\Gamma^{(2)}_{\bar u u,\Lambda}(ip_0)=-\gamma^0(ip_0+\mu)+g\langle\sigma\rangle.
\end{eqnarray}
The inverse propagator at scale $k$ is then an integral of the flow equation two-point function from the ultraviolet $k=\Lambda$ down to $k$
\begin{eqnarray}
G^{-1}_{E,k}(ip_0)&=&\Gamma^{(2)}_{\bar u u,k}(ip_0)\\
&=&-\gamma^0(ip_0+\mu)+g\langle\sigma\rangle+\int_\Lambda^k\partial_k{\Gamma}^{(2)}_{\bar{u}u}(ip_0) dk.\nonumber
\end{eqnarray}
Hence, the Euclidean inverse propagator can be written in terms of k-dependent self energy $G^{-1}_{E,k}(ip_0)=\Gamma^{(2)}_{\bar u u,k}(ip_0)=-\gamma^0(ip_0+\mu)+g\langle\sigma\rangle+\Sigma_k(ip_0)$, with the initial condition $\Sigma_\Lambda(ip_0)=0$. The scale-dependent self-energy in Euclidean space is thus 
\begin{eqnarray}
\Sigma_k(ip_0)-\Sigma_\Lambda(ip_0)=\int_\Lambda^k\partial_k{\Gamma}^{(2)}_{\bar{u}u}(ip_0) dk.
\end{eqnarray}
At ultraviolet limit $k=\Lambda$, no fluctuation is included, $\Sigma_\Lambda(ip_0)=0$ agrees with the bare propagator. As the scale is lowered, quantum fluctuation are included, contributing to the self-energy of the quark. The spectral function is a real-time quantity which requires analytical continuation to bring imaginary time to real time
\begin{eqnarray}
G^{-1}_{R}(\omega)=-G^{-1}_{E}(ip_0\rightarrow\omega+i\eta).
\end{eqnarray}
The inverse retarded propagator is $G^{-1}_{R,k}(\omega)=\gamma^0(\omega+\mu+i\eta)-g\langle\sigma\rangle-\Sigma_{R,k}(\omega)$ and the corresponding retarded self-energy is 
\begin{eqnarray}
\Sigma_{R,k}(\omega)~=~\int_\Lambda^k\partial_k{\Gamma}^{(2)}_{\bar{u}u}(\omega+i\eta)dk.
\end{eqnarray}
The quark propagator at zero momentum $\vec{p}=0$ can be decomposed to the positive energy part and negative energy part 
\begin{eqnarray}
G_{R,k}(\omega)~=~G_{+,k}(\omega)\Lambda_+\gamma^0+G_{-,k}(\omega)\Lambda_-\gamma^0,
\end{eqnarray}
with projection operators $\Lambda_\pm\equiv (1\pm\gamma^0)/2$ acting onto spinors whose chirality is equal(+) or opposite(-) to the helicity. We call +(-)-sector as 'quark' ('anti-quark') sector. Hence, propagator for positive and negative energy parts are
\begin{eqnarray}
G_{\pm,k}(\omega)
&=&\frac{1}{2}\text{Tr}\big[G_{R,k}(\omega)\gamma^0\Lambda_\pm\big]\nonumber\\
&=&\big[\omega+i\eta+\mu\mp m_f-\Sigma_{\pm,k}(\omega)\big]^{-1},
\end{eqnarray}
with 
\begin{eqnarray}
\Sigma_{\pm,k}(\omega)
&=&\frac{1}{2}\text{Tr}\big[\Sigma_{R,k}(\omega)\gamma^0\Lambda_\pm\big].
\end{eqnarray}
We here focus on the self-energy of the 'quark' sector, the flow equation is given by 
\begin{eqnarray}
\partial_k \Sigma_{\pm,k}(\omega)=\frac{1}{2}\text{Tr}\Big[\partial_k{\Gamma}^{(2)}_{\bar{u}u}(\omega)\gamma^0\Lambda_\pm\Big],
\end{eqnarray}
which, after integral over the three momentum, Matsubara sum and analytical continuation, has the following structure,
\begin{eqnarray}
\partial_k\Sigma_{+,k}(\omega)
=g^2\Big(J^S_{\psi\sigma}(\omega)+J^S_{\sigma\psi}(\omega)+3J^{PS}_{\psi\pi}(\omega)+3J^{PS}_{\pi\psi}(\omega)\Big),\nonumber\\
\end{eqnarray}
with $J^S_{\psi\sigma}, J^S_{\sigma\psi}, J^{PS}_{\psi\pi}, J^{PS}_{\pi\psi}$ are the threshold functions presented in the appendix. At vanishing quark number, from the charge conjugation symmetry, we have a relation between $G_\pm$, 
\begin{eqnarray}
G_+(\omega)=-G_-^*(-\omega), 
\end{eqnarray}
but finite chemical potential breaks the charge conjugation symmetry. We limit our study to the case with $\mu=0$ and focus on the spectral function of quark sector only. The spectral function is defined through the imaginary part of the retarded propagator $\rho_k(\omega)=-(1/\pi)\text{Im} G_{R,k}(\omega)$. This can decomposed similarly, $\rho_k(\omega)=\rho_{+,k}(\omega)\Lambda_+\gamma_0+\rho_{-,k}(\omega)\Lambda_-\gamma_0 $, with
\begin{eqnarray}
\rho_{\pm,k}(\omega)
&=&-\frac{1}{\pi}\text{Im} G_{\pm,k}\\
&=&-\frac{1}{\pi}\frac{\text{Im}\Sigma_{\pm,k}(\omega)}{\big(\omega\mp m_f-\text{Re}\Sigma_{\pm,k}(\omega)\big)^2+\text{Im}\Sigma_{\pm,k}(\omega)^2}.\nonumber
\end{eqnarray}

\subsection{Truncation B}
In another truncation, we first evolve the flow equation of the effective potential, and find the infrared quark mass $m_{f,k=0}(T,\mu)$ and meson mass $m_{\sigma,k=0}(T,\mu),~m_{\pi,k=0}(T,\mu)$ at certain temperature and density. We then take these quantities as input to calculate the self-energy of quark at one-loop order. This method has been discussed in Ref. \cite{Kitazawa:2005mp,Kitazawa:2007ep,Kitazawa:2006zi}, here we take the FRG result as an input. In Euclidean space, the fermion self-energy
\begin{eqnarray}
\Sigma_E(ip_0)=&-&g^2\int_q\big[G_\psi(q)G_\sigma(p+q)\nonumber\\
&+&3G_\psi(q)(i\gamma^5)G_\pi(p+q)(i\gamma^5)\big].
\end{eqnarray}
The analytical continuation is taken by $\Sigma_R(\omega)=\Sigma_E(ip_0)\big|_{ip_0\rightarrow \omega+i\eta}$. Taking the energy projection as before, we have the self-energy of 'quark' and 'anti-quark' sector $\Sigma_{\pm}(\omega)
~=~\frac{1}{2}\text{Tr}[\Sigma_{R}(\omega)\gamma^0\Lambda_\pm]$. The self-energy has contribution from the scalar channel and the pseudo-scalar channel, where the interaction mediated by sigma meson and pion respectively,
\begin{eqnarray}
\Sigma_{\pm}(\omega)=\Sigma^S_{\pm}(\omega)+3\Sigma^{PS}_{\pm}(\omega).
\end{eqnarray}
One may first deal with the imaginary part of the scalar channel 
\begin{eqnarray}
\label{ImB}
&&\text{Im}\Sigma^S_+(\omega)~~=~~-\frac{g^2}{2\pi}\int_0^{+\infty}\frac{p^2dp}{E_\sigma}\Big\{\\
&+&\delta(\omega+E_\sigma+E_f)~\Big(1+\frac{m_f}{E_f}\Big)~\Big[1+n_b(E_\sigma)-n_f(E_f)\Big]\nonumber\\
&+&\delta(\omega+E_\sigma-E_f)~\Big(1-\frac{m_f}{E_f}\Big)~\Big[n_b(E_\sigma)+n_f(E_f)\Big]\nonumber\\
&+&\delta(\omega-E_\sigma+E_f)~\Big(1+\frac{m_f}{E_f}\Big)~\Big[n_b(E_\sigma)+n_f(E_f)\Big]\nonumber\\
&+&\delta(\omega-E_\sigma-E_f)~\Big(1-\frac{m_f}{E_f}\Big)~\Big[1+n_b(E_\sigma)-n_f(E_f)\Big]\Big\}.\nonumber
\end{eqnarray}
The momentum integral can be carried out analytically, giving \cite{Kitazawa:2007ep}
\begin{eqnarray}
\text{Im}\Sigma^S_+(\omega)=&-&\frac{g^2}{64\pi}\frac{(\omega+M_+)(\omega-M_-)}{\omega^3}\nonumber\\
&\times&\sqrt{(\omega^2-M_+^2)(\omega^2-M_-^2)}\nonumber\\
&\times&\left[\coth\frac{\omega^2+M_+M_-}{4\omega T}+\tanh\frac{\omega^2-M_-M_+}{4\omega T}\right]\nonumber\\
&\times&\left(\Theta(\omega^2-M_+^2)-\Theta(M_+^2-\omega^2)\right),
\end{eqnarray}
where $M_+=m_\sigma+m_f$ and $M_-=|m_\sigma-m_f|$ and $\Theta(x)$ is the step function. The self-energy of the pseudo-scalar channel can be obtain from that of  the scalar channel by taking the substitution $m_f\rightarrow -m_f$ and $m_\sigma\rightarrow m_\pi$. The self-energy $\Sigma^R(\omega)$ has an ultraviolet divergence which originates from the T-independent part $\Sigma^R_{T=0}(\omega)\equiv\lim_{T\rightarrow 0^+}\Sigma^R(\omega)$. The divergence can be removed by imposing the on-shell renormalization condition on the T-independent part of the quark propagator. The T-dependent part, $\Sigma^R_{T\neq0}(\omega)=\Sigma^R(\omega)-\Sigma^R_{T=0}(\omega)$ is free from divergence. The real part and imaginary part are related by the Krames-Kronig relation, 
\begin{eqnarray}
\text{Re} \Sigma_{R,T\neq0}(\omega)=\mathcal{P}\int_{-\infty}^{+\infty}\frac{dz}{\pi}\frac{\text{Im} \Sigma_{R,T\neq0}(z) }{z-\omega}.
\end{eqnarray}
The spectral function of quark and anti-quark are then given by 
\begin{eqnarray}
\rho_{\pm}(\omega)
&=&-\frac{1}{\pi}\text{Im} G_{\pm}\\
&=&-\frac{1}{\pi}\frac{\text{Im}\Sigma_{\pm}(\omega)}{\big(\omega\mp m_f-\text{Re}\Sigma_{\pm}(\omega)\big)^2+\text{Im}\Sigma_{\pm}(\omega)^2}.\nonumber
\end{eqnarray}

\section{Numerical method and Result}
\label{s3}

\begin{figure}[!hbt]\centering
\includegraphics[height=0.26\textwidth]{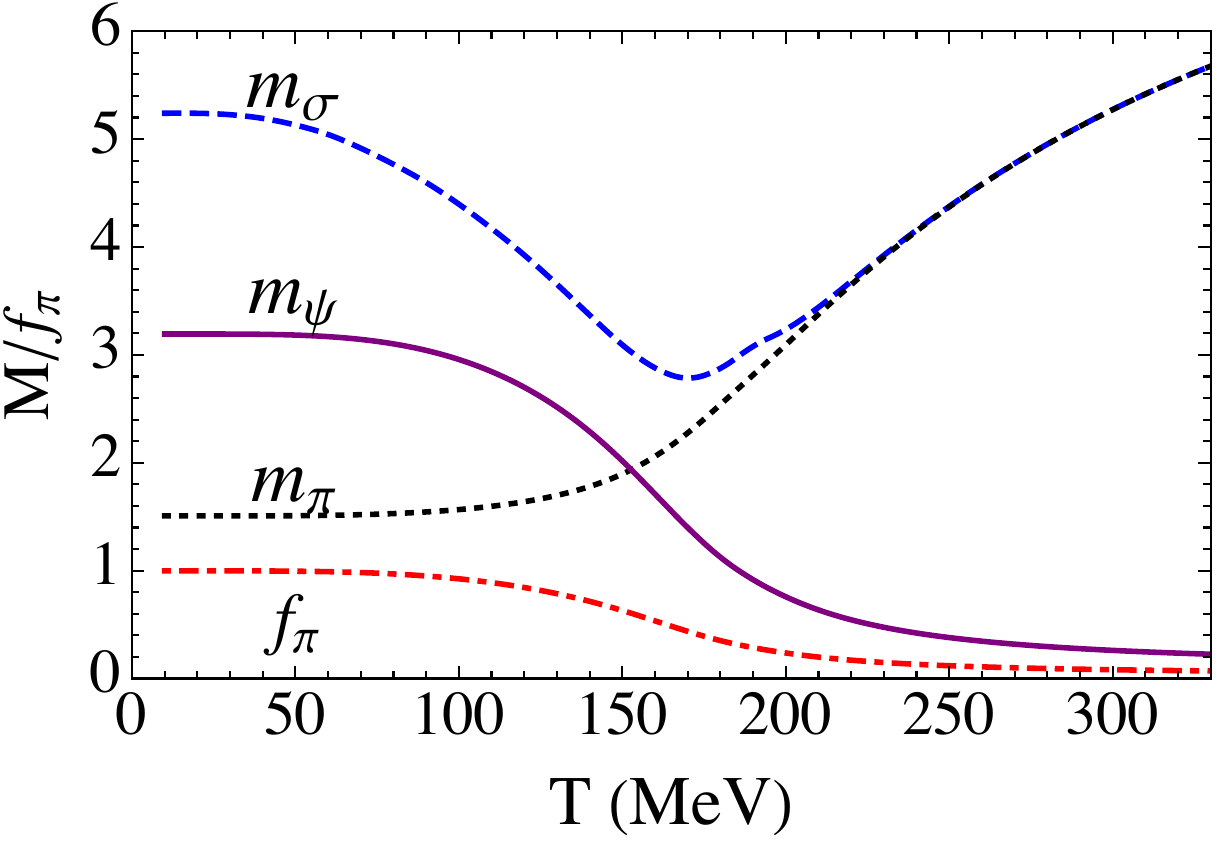}
\caption{Temperature dependence of chiral condensate and various masses.}
\label{fig2}
\end{figure}

To numerically solve the flow equations for the effective potential and the two-point functions, we adopt the grid method, and assume the initial condition at the ultraviolet limit, 
\begin{eqnarray}
\label{ni}
U_\Lambda(\rho)=\frac{1}{2}m_\Lambda^2\rho+\frac{1}{4}\lambda_\Lambda\rho^2,
\end{eqnarray}
for one-dimensional grid, and for the quark self-energy 
\begin{eqnarray}
\Sigma_{+,\Lambda}(\omega)=0.
\end{eqnarray}
During the process of integrating the flow equations from the ultraviolet limit to the infrared limit, the condensates $\langle\sigma\rangle_k$ is obtained by locating the minimum of the $k$-dependent effective potential $U_k$. Choosing the quark mass $m_q=300$ MeV, pion mass $m_\pi=135$ MeV and pion decay constant $f_\pi=93$ MeV in vacuum, the corresponding initial parameters are $m_\Lambda^2/\Lambda^2=0.618$, $\lambda_\Lambda=1$ and $c/\Lambda^3=0.0025$ with the cutoff $\Lambda=900$ MeV. In the numerical calculation, the infrared limit $k=0$ cannot be reached. Instead, the evolution of the flow equation is stopped at $k_{\text{IR}}<10$MeV, where the condensate and coupling have reached a stable structure. By solving the flow equation of the effective potential Eq.(\ref{flowU}), we obtain the dependence of chiral condensate and various masses on temperature, with critical temperature $T_c\sim 170$MeV. The result is presented in Fig.\ref{fig2}.  

The imaginary part of the self-energy is proportional to the difference between the decay and creation rates of the quasi-quark. Each of the four channels has a Dirac delta function, indicating the energy threshold for each process, see Eq.(\ref{ImB},\ref{A5},\ref{A6},\ref{A7}). One can give a physical interpretation to each of the channels, (I) $\delta(\omega+E_\phi+E_\psi)$ describes a annihilation of a quasi-quark with an on-shell free anti-quark and boson, (IV) $\delta(\omega-E_\phi-E_\psi)$ is a decay of a quasi-quark to on-shell free quark and a boson, (II) $\delta(\omega+E_\phi-E_\psi)$ is the decay process of a quasi-quark state Q, into an on-shell quark via a collision with a thermally excited boson, and its inverse process $Q+b\leftrightarrow q$. (III) $\delta(\omega-E_\phi+E_\psi)$ corresponds to a pair annihilation process between the quasi-quark and a thermally excited anti-quark with an emission of a boson into the thermal bath, and its inverse process $Q+\bar{q}\leftrightarrow b$. The latter two processes are called Landau damping, which vanishes at $T=0$, as it involves thermally excited particles in the initial states. The Landau damping plays an important role in the spectral function as temperature rises and is closely related to the three peak structure at high temperature. Process (II) and (III) both cause a mixing between the quark and anti-quark hole through coupling to thermally excited boson as discussed in Ref. \cite{Kitazawa:2007ep}.

In truncation A, the real part and imaginary part of the self-energy are calculated separately, the real part is given by the principle value integral. While for the imaginary part of the two-point function, the integral over RG-scale $k$ only has contribution from a few scales $k_i$ due to the appearance of aforementioned Dirac delta functions. The following structures are encountered in the integration of the imaginary part of the flow equation, where $g(k)=\pm E_{\phi,k}\pm(\mp) E_{\psi,k}$, and $k_i$ are zero points of the delta-function at a certain energy $\omega$, with $\omega+g(k_i)=0$,
\begin{eqnarray}
&&\int_\Lambda^0f(k)\delta(\omega+g(k))dk=-\sum_i\frac{f(k_i)}{|g'(k_i)|},\\
&&\int_\Lambda^0f(k)\delta'(\omega+g(k))dk\nonumber\\
&&\qquad=\sum_i\left(\frac{f'(k_i)}{g'(k_i)}-\frac{f(k_i)g''(k_i)}{g'(k_i)^2}\right)\frac{1}{|g'(k_i)|}.\nonumber
\end{eqnarray}

\begin{figure}[!hbt]\centering
\includegraphics[width=0.45\textwidth]{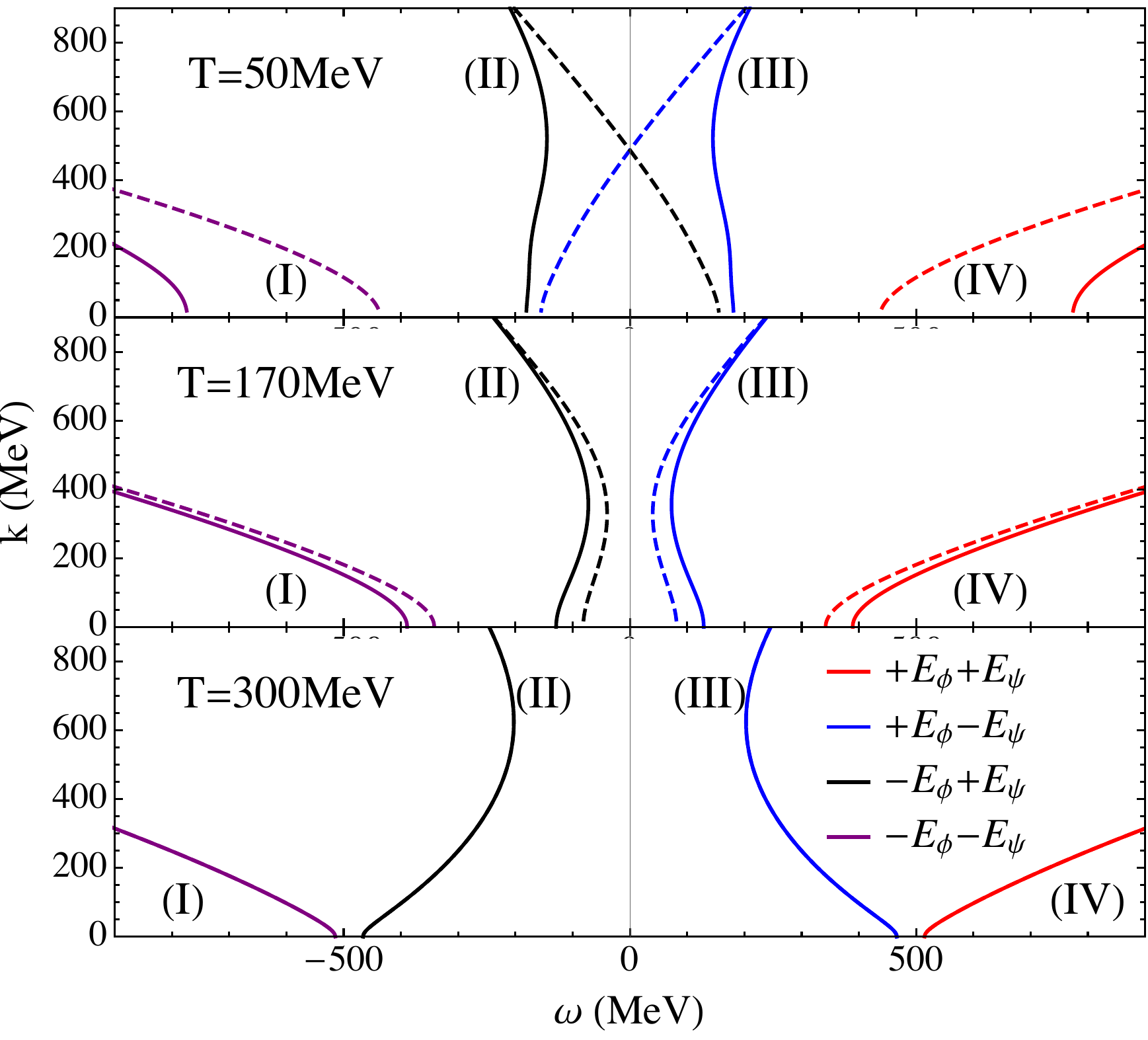}
\caption{The RG-scale dependence of threshold for channels (I) to (IV), $\pm E_{\phi,k}\pm E_{\psi,k}$, ($\phi=\sigma, \pi$) for three temperature $T=50,~170,~300$MeV.}
\label{fig3}
\end{figure}

It is required that $g(k)$ is a continuously differentiable function with $g'$ nowhere zero. In the integral of the flow equation $g(k)$ has certain points where the derivatives are zero, the domain must be broken up to exclude the $g'= 0$ point. These $g'(k_i)= 0$ points are similar to the van-Hove singularities in the density of states in condensed matter physics\cite{VanHove:1953}, $g(\omega)=\sum_n\int\frac{d^3k}{(2\pi)^3}\delta(\omega-\omega_n(\vec k))=\sum_n\int\frac{dS_\omega}{(2\pi)^3}\frac{1}{|\nabla\omega_n(\vec k)|}$. The group velocity $\nabla\omega_n(\vec k)$ vanishes at certain momenta, resulting a divergent integrand. The divergence is integrable in 3-dimensions, in lower dimensions, the van-Hove singularity appears. The van Hove singularities have also attracted interests in high energy physics~\cite{Mustafa:1999dt,Peshier:1999dt,Karsch:2000gi, Mustafa:2002pb}. In our case, the integral over $k$ is one-dimensional and gives divergence in the imaginary part at finite temperature. This divergence exists only in truncation A, when scale dependence of various masses are taken into consideration, and appears only when two conditions are satisfied at the same time, that $\omega+\delta E(k^*)=0$ and $\delta E'(k^*)=0$. The scale dependence of meson mass also causes divergence in the real part of the self-energy. In FIG.\ref{fig3}, we present the RG-scale dependence of threshold for channels (I) to (IV), $\pm E_{\phi,k}\pm E_{\psi,k}$, ($\phi=\sigma, \pi$) for three temperature $T=50,~170,~300$MeV, with the dashed lines for pions and solid line for sigma. For terms related to Landau damping (II) and (III), at certain $\omega$, the k-integral runs into points that  $\delta E'(k^*)=0$, leading to the divergence in imaginary part. At low temperature, when the Landau damping is suppressed, the diverge does not appear. In contrast, in truncation B, we take $m_{\phi,k=0}$ as input, the imaginary part and real part are always finite after remove the zero-temperature part. 

\begin{figure}[!hbt]\centering
\includegraphics[width=0.42\textwidth]{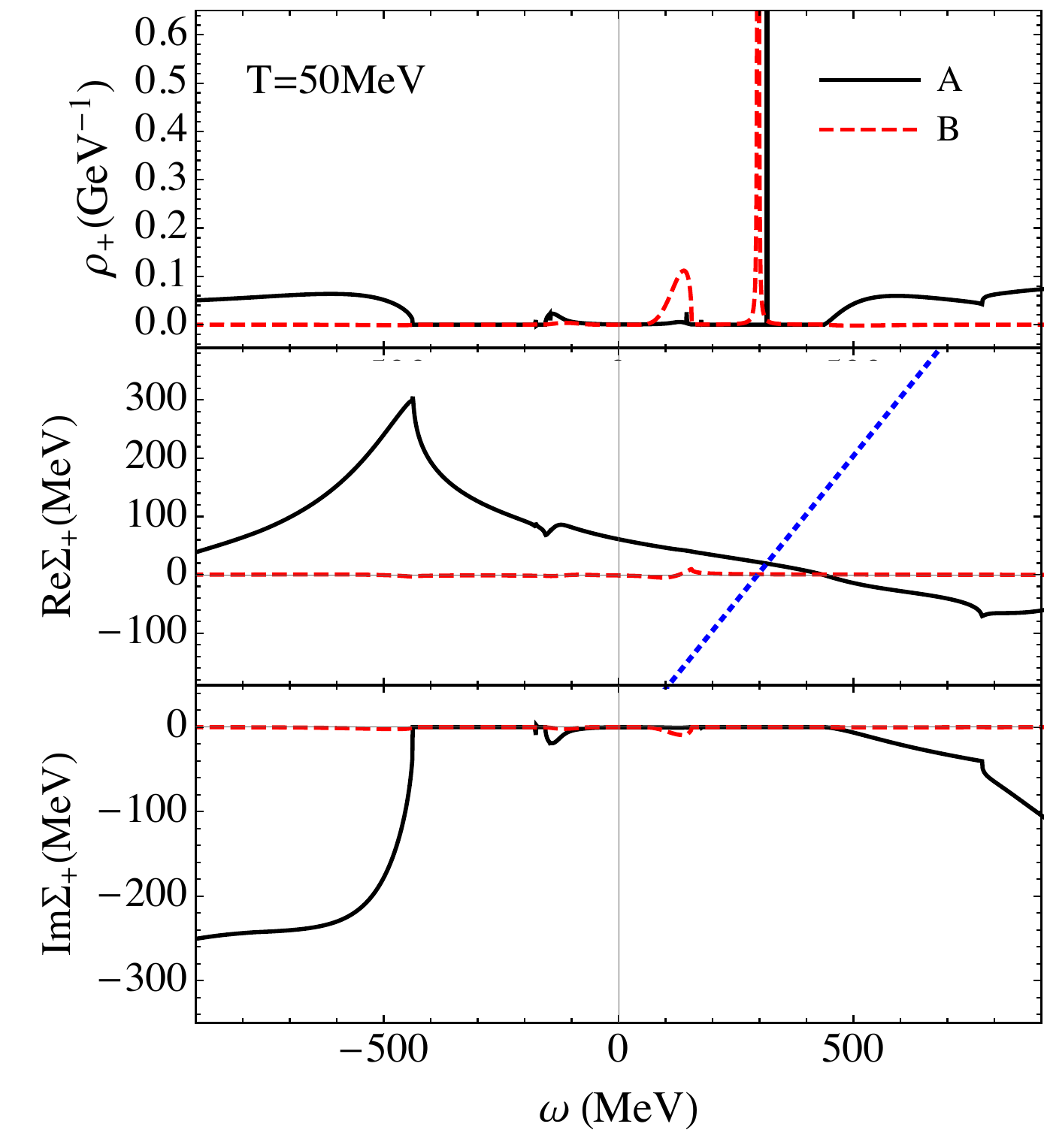}
\caption{Quark spectral function at $T=50$MeV, from top to bottom are spectral function, real and imaginary part of the quark self-energy. The black solid line stands for truncation A, while red dashed line for truncation B. The blue dotted line represent $\omega-m_f$, with $m_f=296$MeV. }
\label{fig4}
\end{figure}

We first present the spectral function of quark sector at relatively low temperature, In FIG.\ref{fig4}, from top to bottom, are figures for the spectral function, the real part and imaginary part of the self-energy, with the black solid line represents truncation A, and red dashed line for truncation B. The zero-temperature result can be found in Ref. \cite{Tripolt:2018qvi}, to which, we have also found the same result. At T=50 MeV, the system is still in the chiral symmetry breaking phase, with quark mass $m_f=296$MeV and meson mass $m_\sigma=477$MeV and $m_\pi=140$MeV. In both truncations we have delta-peaks around fermion mass, $\omega=315$MeV for truncation A, and $\omega=296$MeV for truncation B. The peak structure in $\rho_+(\omega)$ emerges at the "quasi-pole" \cite{Kitazawa:2007ep} which is defined as zero of the real part of the inverse propagator $\omega-m_f-\text{Re}\Sigma_+(\omega)=0$, providing that the imaginary part is small enough at that point. In both truncations, the real part of the inverse propagator only has one "quasi-pole", which is the cross point of $\omega-m_f$ (the blue dotted line) with $\text{Re}\Sigma_+(\omega)$ in the figure. The difference in the position of the peak in both truncations comes the inclusion and subtraction of different fluctuation. The Landau damping is still well suppressed, giving small imaginary part and flat structures in the spectral function at low energy. At large energy, when $|\omega|>m_\psi+m_\phi$ the decay processes (I) and (IV) take place, giving finite imaginary part and the continuous spectrum in the spectral function in truncation A. While in truncation B, this continuous spectrum is subtracted when performing the renormalization.

\begin{figure}[!hbt]\centering
\includegraphics[width=0.42\textwidth]{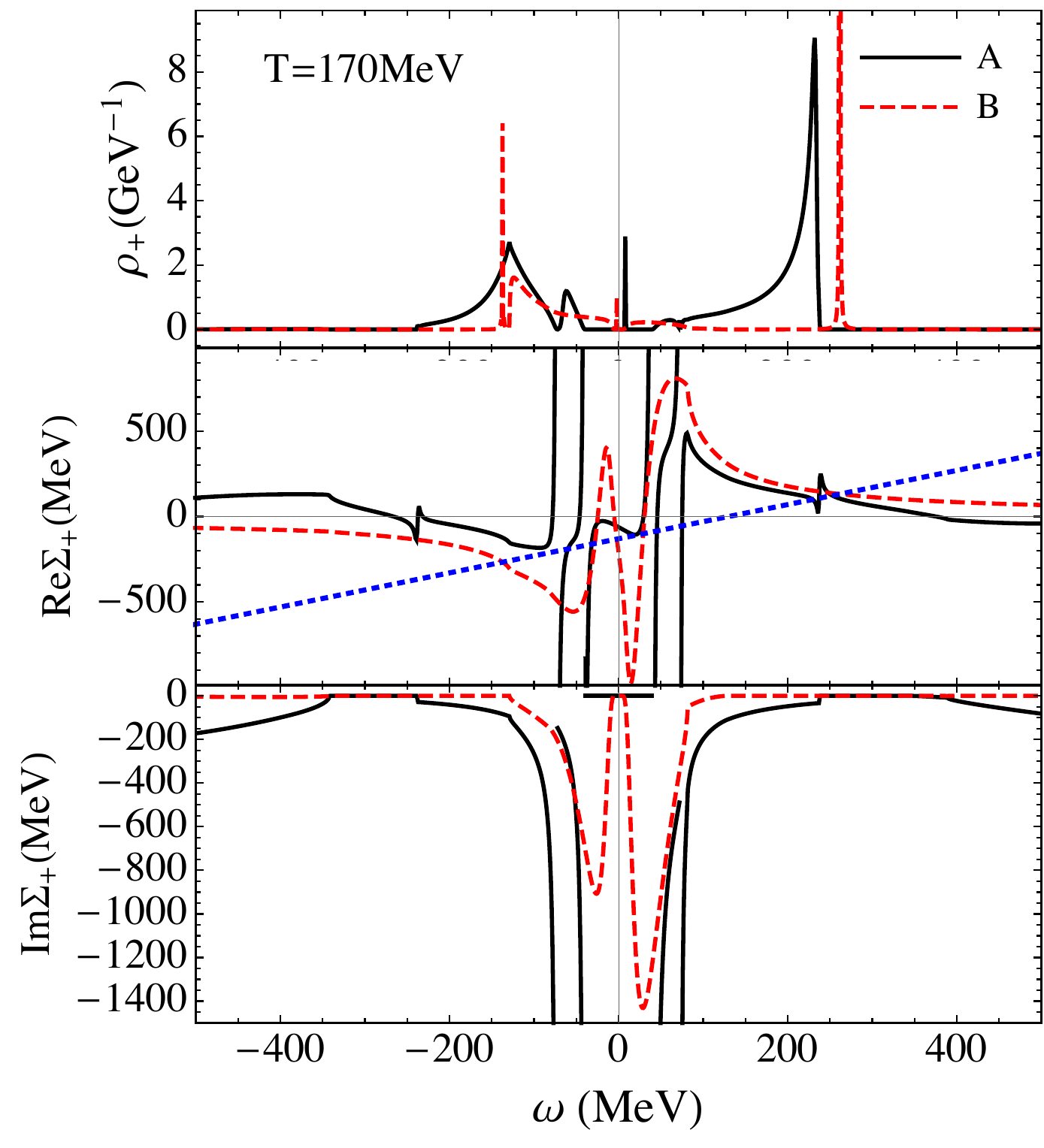}
\caption{Quark spectral function at $T=170$MeV, from top to bottom are spectral function, real and imaginary part of the quark self-energy. The black solid line stands for truncation A, while red dashed line for truncation B. The blue dotted line represent $\omega-m_f$, with $m_f=130$MeV.}
\label{fig5}
\end{figure}

The chiral phase transition takes place at the temperature about $T=170$MeV, pion and sigma meson have not been degenerate yet, with $m_f=130$MeV, $m_\sigma=259$MeV, $m_\pi=212$MeV. For truncation A, the threshold of various channels is presented in FIG.\ref{fig3}. In the scattering channel with the thermally excited boson (II) and the annihilation channel with the thermally excited anti-quark (III), $\pm E_{\phi,k}\mp E_{\psi,k}$ has points where the derivative with RG-scale vanished. This brings about divergence in the imaginary part, and oscillation in the real part, see black lines in Fig.\ref{fig5}. The imaginary part is discontinuous at the four van-Hove singularities, and is zero when $\omega\leq|\pm E_{\phi,k}\mp E_{\psi,k}|_{\text{min}}$, where process (II) (III) are forbidden. $\omega-m_f-\text{Re}\Sigma_+(\omega)$ has several quasi-poles in truncation A, yet the spectral function has only one peak, when the imaginary part is small $\omega=21$MeV indicating a quasi-particle mode here. For other quasi-poles, the imaginary parts are too large to form a peak, giving several bumps instead. In truncation B, however, we have a quite different case, the imaginary part is non-zero but finite when process (II) and (III) are allowed. The real part has five quasi-poles, at three of them, the imaginary part is small enough to give a peak in the spectral function. When process (II) and (III) are allowed, the imaginary part is large and gives only small bumps in the spectral function, see the red dashed lines in FIG.\ref{fig5}. The spectral function presents a three peak structure in truncation B, with one peak at the origin, and two quasi-pole where the imaginary part is small. This is the typical structure at $T\sim m_b$ and has been discussed in-depth in Ref.\cite{Kitazawa:2007ep}. The Landau damping, which causes peaks in imaginary part, is essential in the three peak structure in the spectral function. In truncation A, the spectral function also presents a peak at small $\omega$ but not at at the exact origin, namely, instead of a zero mode, we have a soft-mode in truncation A, which also arises from the Landau damping effect. 

Finally, we analyze the spectral function in both truncations at $T=300$MeV, where the system has reached the chiral restored phase, with $m_f=24$MeV, and degenerate meson mass $m_\sigma\approx m_\pi=490$MeV. For truncation A, the threshold of each channel is presented in the last figure in FIG.\ref{fig3}. There is a large area where channel (II) (III) are forbidden, leading to $\text{Im}\Sigma_+=0$ at small energy. Channel (II) (III) both have points where $\pm E'_{\phi,k^*}\mp E'_{\psi,k^*}=0$, where the imaginary part diverges. The real part of the inverse propagator has 7 quasi-poles, at two of which, the imaginary parts are too large to give a peak at $\omega=10,\pm 160$MeV. For the three quasi-poles at low energy, the imaginary part is small and gives three peaks. For the two quasi-poles at large $\omega$, the Landau damping effect gives two bumps in the spectral function.  While in truncation B, the Landau damping effect again gives two peaks in the imaginary part of the self-energy, and an oscillation in real part. $\omega-m_f-\text{Re}\Sigma_+(\omega)=0$ has five quasi-poles, $\text{Im}\Sigma_+$ has relative large values at four of the quasi-poles, leading to two peaks with finite width, and a delta-peak at the origin. 
\begin{figure}[!hbt]\centering
\includegraphics[width=0.42\textwidth]{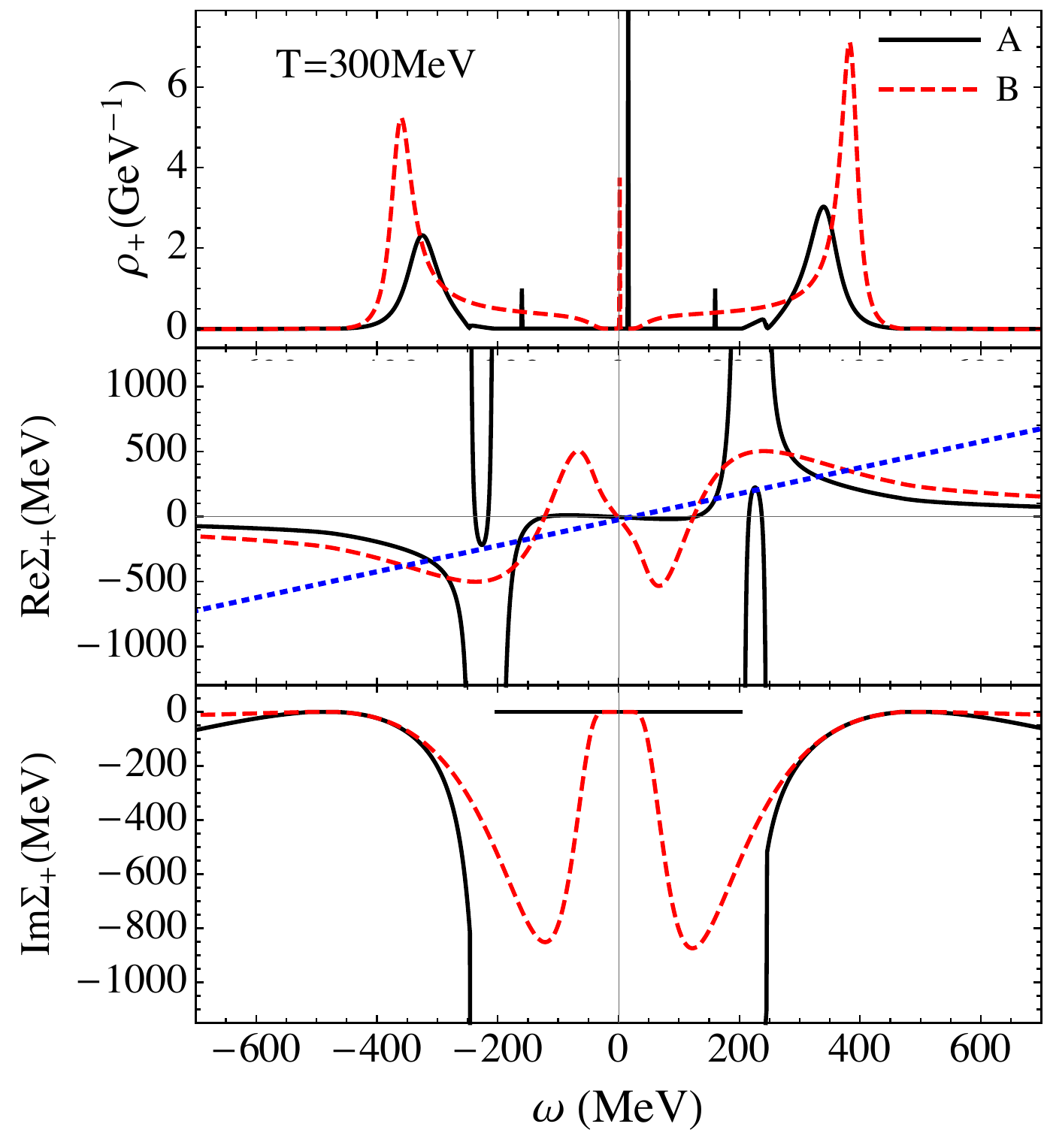}
\caption{Quark spectral function at $T=300$MeV, from top to bottom are spectral function, real and imaginary part of the quark self-energy. The black solid line stands for truncation A, while red dashed line for truncation B. The blue dotted line represent $\omega-m_f$, with $m_f=24$MeV.}
\label{fig6}
\end{figure}

Two scales are of crucial importance in the structure of the fermion spectral function, $m_f/m_b$ and $T/m_b$. The three peak structure is most obvious when $m_f/m_b\ll 1$ and $T/m_b\sim 1$. When these two factors are approached, the peaks become higher and sharper. For the chiral crossover, $T/m_b\sim 1$ takes place around the critical temperature, where we observe three sharp peaks in truncation B. For high temperature $T/m_b\approx0.61$, $m_f/m_b\ll1$ is satisfied, thus fermion is almost massless, the three peak structure is also obvious but with finite width. In truncation B, the appearance of zero-mode can be demonstrated by analyzing the real and imaginary part of self-energy at $\omega=0$. In the case of finite fermion mass, one always has  $\text{Im}\Sigma_+(\omega=0)=0$, $\text{Re}\Sigma_+(\omega)$ can be exactly calculated to give a quasi-pole very close to the origin, thus a peak will appear at the origin. In truncation A, the multi-peak structure is also observed as $T/m_b$ approaches unit. However, the zero-mode becomes a soft-mode, where the quasi-pole is slightly away from the origin. The disappearance of this zero-mode results from the limited scale of momentum in the propagator and the scale dependence of the meson masses. One may expect that with larger $\Lambda$, the soft-mode would be closer to the origin. 

\section{Summary}
\label{s4}
We investigate the spectral function of quark in two truncations in a quark-meson model with functional renormalization group. In both truncations, we first solve the flow equation of the effective potential and find out the temperature and scale dependence of fermion and meson masses. In truncation A, we take the scale-dependent masses in step one as the input, and evolve the flow equation of the two-point function; in truncation B, we take the masses in the infrared as input and calculate the one-loop self-energy. After the analytical continuation, we have the spectral function. 

When one consistently integrates the flow equation of the two-point function, the RG-scale dependence of the energy thresholds of decay and creation channels leads to van Hove singularities at finite temperature, where Landau damping plays an important role. This singularity leads to divergence in both the real and imaginary part. Another feature is that, at high temperature, the Landau damping is forbidden at low energy, leading to zero imaginary part and several peaks in the low energy area. 

In comparison, when directly calculating the one-loop self-energy, one gets a three-peak structure when temperature rises and becomes comparable to meson mass. The spectral function has a peak at the origin, namely a fermion zero-mode, and the other peaks comes from the Landau damping effect. 

Our work presents a first calculation of finite temperature quark spectral function, and supports quasi-particle picture of quarks around the critical temperature.

{\bf Note Added:} During the writing of this manuscript, we became aware of the work by R. A. Tripolt et al. \cite{Tripolt:2018qvi}, where the fermion spectral function at zero temperature was investigated.

{\bf Acknowledgement:} The work is supported by the National Natural Science Foundation of China (Grant Nos. 11335005, 11575093, and 11775123), MOST  (Grant Nos. 2013CB922000 and 2014CB845400), and Tsinghua University Initiative Scientific Research Program.

\begin{appendix}
\section{Loop functions}
\label{threshold}
\begin{widetext}
With the boson and fermion occupation numbers and their derivatives,
\begin{eqnarray}
&& n_B(x)=\frac{1}{e^{x/T}-1},\quad n_F(x)=\frac{1}{e^{x/T}+1},\nonumber\\
&& n'_B(x)=\frac{dn_B(x)}{d x},\quad~~ n'_F(x)=\frac{dn_F(x)}{d x},
\end{eqnarray}
the loop functions $J_{\phi}$ and $J_\psi$ in the flow equation for effective potential are explicitly expressed as
\begin{eqnarray}
J_\phi &=& \frac{k^4}{3\pi^2} \frac{1+2n_B(E_\phi)}{2E_\phi},\nonumber\\
J_\psi &=& \frac{k^4}{3\pi^2} \frac{1-n_F(E_\psi-\mu)-n_F(E_\psi+\mu)}{E_\psi}.
 \label{A2}
\end{eqnarray}
The threshold function $J^S_{\psi\sigma}, J^S_{\sigma\psi}, J^{PS}_{\psi\pi}, J^{PS}_{\pi\psi}$ in the two-point function can be obtained taking derivatives with respect to the corresponding energy, 
\begin{eqnarray}
J^S_{\psi\sigma}(ip_0)&=&-\frac{1}{2E_\psi}\frac{\partial}{\partial E_\psi}J^S(ip_0),\qquad~~
J^S_{\sigma\psi}(ip_0)~=~-\frac{1}{2E_\sigma}\frac{\partial}{\partial E_\sigma}J^S(ip_0),\nonumber\\
J^{PS}_{\psi\pi}(ip_0)&=&-\frac{1}{2E_\psi}\frac{\partial}{\partial E_\psi}J^{PS}(ip_0),\qquad
 J^{PS}_{\pi\psi}(ip_0)~=~-\frac{1}{2E_\pi}\frac{\partial}{\partial E_\pi}J^{PS}(ip_0).
 \label{A3}
\end{eqnarray}
After the analytical continuation, $J^S$ in Minkovski space is 
\begin{eqnarray}
\label{A4}
J^S(\omega)=-\frac{k^4}{3\pi^2}\frac{1}{4E_\phi}
\bigg\{&&\frac{1}{\omega+\mu+E_\phi+E_\psi+i\eta}\left(1-\frac{m_f}{E_\psi}\right)(1+n_B(E_\phi)-n_F(E_\psi+\mu))\nonumber\\
&+&\frac{1}{\omega+\mu+E_\phi-E_\psi+i\eta}\left(1+\frac{m_f}{E_\psi}\right)(n_B(E_\phi)+n_F(E_\psi-\mu))\nonumber\\
&+&\frac{1}{\omega+\mu-E_\phi+E_\psi+i\eta}\left(1-\frac{m_f}{E_\psi}\right)(n_B(E_\phi)+n_F(E_\psi+\mu))\nonumber\\
&+&\frac{1}{\omega+\mu-E_\phi-E_\psi+i\eta}\left(1+\frac{m_f}{E_\psi}\right)(1+n_B(E_\phi)-n_F(E_\psi-\mu))\bigg\},
\end{eqnarray}
Making substitution by $m_f\rightarrow-m_f$ and $E_\sigma\rightarrow E_\pi$ then we can get threshold for pseudoscalar channel $J^{PS}(\omega)$. The real part is given by the principle value, while the imaginary part of the threshold function is then, 
\begin{eqnarray}
 \label{A5}
\text{Im}J^S(\omega)~=~\frac{k^4}{3\pi}\frac{1}{4E_\phi}~\Big\{& &\delta(\omega+\mu+E_\phi+E_\psi)\left(1-\frac{m_f}{E_\psi}\right)\Big(1+n_B(E_\phi)-n_F(E_\psi+\mu)\Big)\nonumber\\
&+&\delta(\omega+\mu+E_\phi-E_\psi)\left(1+\frac{m_f}{E_\psi}\right)\Big(n_B(E_\phi)+n_F(E_\psi-\mu)\Big)\nonumber\\
&+&\delta(\omega+\mu-E_\phi+E_\psi)\left(1-\frac{m_f}{E_\psi}\right)\Big(n_B(E_\phi)+n_F(E_\psi+\mu)\Big)\nonumber\\
&+&\delta(\omega+\mu-E_\phi-E_\psi)\left(1+\frac{m_f}{E_\psi}\right)\Big(1+n_B(E_\phi)-n_F(E_\psi-\mu)\Big)\Big\}.
\end{eqnarray}
Following Eq.(\ref{A3}), one have the imaginary part of the threshold functions $\text{Im}J^S_{\psi\phi}$, $\text{Im}J^S_{\phi\psi}$ for the scalar channel.
\begin{eqnarray}
 \label{A6}
\text{Im}J^S_{\psi\phi}(\omega)
~=~-\frac{k^4}{3\pi}\frac{1}{8E_\phi E_\psi}~\bigg\{&&\delta'(\omega+\mu+E_\phi+E_\psi)\left(1-\frac{m_f}{E_\psi}\right)\Big(1+n_B(E_\phi)-n_F(E_\psi+\mu)\Big)\nonumber\\
&-&\delta'(\omega+\mu+E_\phi-E_\psi)\left(1+\frac{m_f}{E_\psi}\right)\Big(n_B(E_\phi)+n_F(E_\psi-\mu)\Big)\nonumber\\
&+&\delta'(\omega+\mu-E_\phi+E_\psi)\left(1-\frac{m_f}{E_\psi}\right)\Big(n_B(E_\phi)+n_F(E_\psi+\mu)\Big)\nonumber\\
&-&\delta'(\omega+\mu-E_\phi-E_\psi)\left(1+\frac{m_f}{E_\psi}\right)\Big(1+n_B(E_\phi)-n_F(E_\psi-\mu)\Big)\bigg\}\\
-\frac{k^4}{3\pi}\frac{m_f}{8E_\phi E_\psi^3}~\bigg\{&&\delta(\omega+\mu+E_\phi+E_\psi)\left[1+n_B(E_\phi)-n_F(E_\psi+\mu)-E_\psi\left(\frac{E_\psi}{m_f}-1\right)n'_F(E_\psi+\mu)\right]\nonumber\\
&-&\delta(\omega+\mu+E_\phi-E_\psi)\left[n_B(E_\phi)+n_F(E_\psi-\mu)-E_\psi\left(\frac{E_\psi}{m_f}+1\right)n'_F(E_\psi-\mu)\right]\nonumber\\
&+&\delta(\omega+\mu-E_\phi+E_\psi)\left[n_B(E_\phi)+n_F(E_\psi+\mu)+E_\psi\left(\frac{E_\psi}{m_f}-1\right)n'_F(E_\psi+\mu)\right]\nonumber\\
&-&\delta(\omega+\mu-E_\phi-E_\psi)\left[1+n_B(E_\phi)-n_F(E_\psi-\mu)+E_\psi\left(\frac{E_\psi}{m_f}+1\right)n'_F(E_\psi-\mu)\right]\bigg\}\nonumber
\end{eqnarray}
\begin{eqnarray}
 \label{A7}
\text{Im}J^S_{\phi\psi}(\omega)
~=~-\frac{k^4}{3\pi}\frac{1}{8E_\phi^2}~\Big\{&&\delta'(\omega+\mu+E_\phi+E_\psi)\left(1+\frac{m_f}{E_\psi}\right)\Big(1+n_B(E_\phi)-n_F(E_\psi+\mu)\Big)\nonumber\\
&+&\delta'(\omega+\mu+E_\phi-E_\psi)\left(1-\frac{m_f}{E_\psi}\right)\Big(n_B(E_\phi)+n_F(E_\psi-\mu)\Big)\nonumber\\
&-&\delta'(\omega+\mu-E_\phi+E_\psi)\left(1+\frac{m_f}{E_\psi}\right)\Big(n_B(E_\phi)+n_F(E_\psi+\mu)\Big)\nonumber\\
&-&\delta'(\omega+\mu-E_\phi-E_\psi)\left(1-\frac{m_f}{E_\psi}\right)\Big(1+n_B(E_\phi)-n_F(E_\psi-\mu)\Big)\Big\}\\
+\frac{k^4}{3\pi}\frac{1}{8E_\phi^3}~\Big\{&&\delta(\omega+\mu+E_\phi+E_\psi)\left(1+\frac{m_f}{E_\psi}\right)\Big(1+n_B(E_\phi)-n_F(E_\psi+\mu)-E_\phi n'_B(E_\phi)\Big)\nonumber\\
&+&\delta(\omega+\mu+E_\phi-E_\psi)\left(1-\frac{m_f}{E_\psi}\right)\Big(n_B(E_\phi)+n_F(E_\psi-\mu)-E_\phi n'_B(E_\phi)\Big)\nonumber\\
&+&\delta(\omega+\mu-E_\phi+E_\psi)\left(1+\frac{m_f}{E_\psi}\right)\Big(n_B(E_\phi)+n_F(E_\psi+\mu)-E_\phi n'_B(E_\phi)\Big)\nonumber\\
&+&\delta(\omega+\mu-E_\phi-E_\psi)\left(1-\frac{m_f}{E_\psi}\right)\Big(1+n_B(E_\phi)-n_F(E_\psi-\mu)-E_\phi n'_B(E_\phi)\Big)\Big\}\nonumber
\end{eqnarray}
\end{widetext}
\end{appendix}

\bibliographystyle{unsrt}

\begin{thebibliography}{10}

\bibitem{Adams:2005dq} 
  J.~Adams {\it et al.} [STAR Collaboration],
  Nucl.\ Phys.\ A {\bf 757}, 102 (2005).

\bibitem{Adcox:2004mh} 
  K.~Adcox {\it et al.} [PHENIX Collaboration],
  Nucl.\ Phys.\ A {\bf 757}, 184 (2005).

\bibitem{Karsch:2007wc} 
  F.~Karsch and M.~Kitazawa,
  Phys.\ Lett.\ B {\bf 658}, 45 (2007).

\bibitem{Karsch:2009tp} 
  F.~Karsch and M.~Kitazawa,
  Phys.\ Rev.\ D {\bf 80}, 056001 (2009).

\bibitem{Kaczmarek:2012mb} 
  O.~Kaczmarek, F.~Karsch, M.~Kitazawa and W.~Soldner,
  Phys.\ Rev.\ D {\bf 86}, 036006 (2012).
  
 \bibitem{Harada:2008vk} 
  M.~Harada and Y.~Nemoto,
  Phys.\ Rev.\ D {\bf 78}, 014004 (2008)
 
 \bibitem{Mueller:2010ah} 
  J.~A.~Mueller, C.~S.~Fischer and D.~Nickel,
  Eur.\ Phys.\ J.\ C {\bf 70}, 1037 (2010).

\bibitem{Pisarski:1988vd} 
  R.~D.~Pisarski,
  Phys.\ Rev.\ Lett.\  {\bf 63}, 1129 (1989).
  
\bibitem{Braaten:1989mz} 
  E.~Braaten and R.~D.~Pisarski,
  Nucl.\ Phys.\ B {\bf 337}, 569 (1990).
  
\bibitem{Su:2014rma} 
  N.~Su and K.~Tywoniuk,
  Phys.\ Rev.\ Lett.\  {\bf 114}, no. 16, 161601 (2015).

\bibitem{Kitazawa:2005mp} 
  M.~Kitazawa, T.~Kunihiro and Y.~Nemoto,
  Phys.\ Lett.\ B {\bf 633}, 269 (2006).
  
\bibitem{Kitazawa:2007ep} 
  M.~Kitazawa, T.~Kunihiro, K.~Mitsutani and Y.~Nemoto,
  Phys.\ Rev.\ D {\bf 77}, 045034 (2008).  

\bibitem{Kitazawa:2006zi} 
  M.~Kitazawa, T.~Kunihiro and Y.~Nemoto,
  Prog.\ Theor.\ Phys.\  {\bf 117}, 103 (2007).

\bibitem{Berges:2000ew}
  J.~Berges, N.~Tetradis and C.~Wetterich,
  Phys.\ Rept.\  {\bf 363}, 223 (2002).

\bibitem{Polonyi:2001se}
  J.~Polonyi,
  Central Eur.\ J.\ Phys.\  {\bf 1}, 1 (2003).

\bibitem{Pawlowski:2005xe}
  J.~M.~Pawlowski,
  Annals Phys.\  {\bf 322}, 2831 (2007).

\bibitem{Gies:2006wv}
  H.~Gies,
  Lect.\ Notes Phys.\  {\bf 852}, 287 (2012).

\bibitem{Kopietz:2010zz}
  P.~Kopietz, L.~Bartosch and F.~Schütz,
  Lect.\ Notes Phys.\  {\bf 798}, 1 (2010).

\bibitem{Braun:2011pp}
  J.~Braun,
  J.\ Phys.\ G {\bf 39}, 033001 (2012).

\bibitem{Kamikado:2013sia}
  K.~Kamikado, N.~Strodthoff, L.~von Smekal and J.~Wambach,
  Eur.\ Phys.\ J.\ C {\bf 74}, no. 3, 2806 (2014).

\bibitem{Pawlowski:2015mia}
  J.~M.~Pawlowski and N.~Strodthoff,
  Phys.\ Rev.\ D {\bf 92}, no. 9, 094009 (2015).

  \bibitem{Strodthoff:2016pxx} 
  N.~Strodthoff,
  Phys.\ Rev.\ D {\bf 95}, no. 7, 076002 (2017).

\bibitem{Tripolt:2013jra}
  R.~A.~Tripolt, N.~Strodthoff, L.~von Smekal and J.~Wambach,
  Phys.\ Rev.\ D {\bf 89}, no. 3, 034010 (2014).

\bibitem{Tripolt:2014wra}
  R.~A.~Tripolt, L.~von Smekal and J.~Wambach,
  Phys.\ Rev.\ D {\bf 90}, no. 7, 074031 (2014).
  
  \bibitem{Jung:2016yxl} 
  C.~Jung, F.~Rennecke, R.~A.~Tripolt, L.~von Smekal and J.~Wambach,
  Phys.\ Rev.\ D {\bf 95}, no. 3, 036020 (2017).

\bibitem{Tripolt:2016cey} 
  R.~A.~Tripolt, L.~von Smekal and J.~Wambach,
  Int.\ J.\ Mod.\ Phys.\ E {\bf 26}, no. 01n02, 1740028 (2017).

\bibitem{Yokota:2016kyz}
  T.~Yokota, T.~Kunihiro and K.~Morita,
  PTEP {\bf 2016}, no. 7, 073D01 (2016).

\bibitem{Tripolt:2018qvi} 
  R.~A.~Tripolt, J.~Weyrich, L.~von Smekal and J.~Wambach,
  arXiv:1807.11708 [hep-ph].

\bibitem{Wang:2017vis} 
  Z.~Wang and P.~Zhuang,
  Phys.\ Rev.\ D {\bf 96}, no. 1, 014006 (2017).

\bibitem{Jungnickel:1995fp}
  D.~U.~Jungnickel and C.~Wetterich,
  Phys.\ Rev.\ D {\bf 53}, 5142 (1996).

\bibitem{Schaefer:2004en}
  B.~J.~Schaefer and J.~Wambach,
  Nucl.\ Phys.\ A {\bf 757}, 479 (2005).
  
\bibitem{Pawlowski:2014zaa} 
  J.~M.~Pawlowski and F.~Rennecke,
  Phys.\ Rev.\ D {\bf 90}, no. 7, 076002 (2014).

\bibitem{Litim:2000}
  D.~F.~Litim,
  Phys.\ Lett.\ B {\bf 486}, 92 (2000).

\bibitem{Litim:2001up}
  D.~F.~Litim,
  Phys.\ Rev.\ D {\bf 64}, 105007 (2001).

\bibitem{VanHove:1953}
L.~Van~Hove.
Phys.\ Rev.{\bf 89} (1953) 1189.

\bibitem{Mustafa:1999dt} 
  M.~G.~Mustafa, A.~Schafer and M.~H.~Thoma,
  Phys.\ Rev.\ C {\bf 61}, 024902 (2000).

\bibitem{Peshier:1999dt}
  A.~Peshier and M.~H.~Thoma,
  Phys.\ Rev.\ Lett.\  {\bf 84}, 841 (2000).
  
\bibitem{Karsch:2000gi} 
  F.~Karsch, M.~G.~Mustafa and M.~H.~Thoma,
  Phys.\ Lett.\ B {\bf 497}, 249 (2001).
  
\bibitem{Mustafa:2002pb} 
  M.~G.~Mustafa and M.~H.~Thoma,
  Pramana {\bf 60}, 711 (2003).










\end{thebibliography}

\end{document}